\documentstyle[11pt]{article}
\begin{document}
\title{\bf Critical Temperature of a Trapped Interacting Bose Gas\\ 
in the Local Density Approximation}
\author{\it Hualin Shi and Wei-Mou Zheng\\
\it Institute of Theoretical Physics, Academia Sinica, Beijing 100080, China}
\date{}
\maketitle
\centerline{\bf Abstract}
\smallskip
\begin{center}
\parbox{14cm}{The Bose gas in an external potential is studied by means of the
local density approximation. An analytical result is derived for the 
dependence of the critical temperature of Bose-Einstein condensation on 
the mutual interaction in a generic power-law potential.}
\end{center}

\leftline{PACS numbers: 03.75.Fi, 32.80.Pj}

In an ideal Bose-Einstein gas, the zero momentum state can
become macroscopically occupied, and the system then undergoes
a phase transition --- the Bose-Einstein condensation (BEC)
\cite{huang87}. The $\lambda$-transition of liquid helium-II is
often interpreted essentially as a result of BEC, the strongest
degeneracy effect of a Boson system\cite{pen56}. For many years,
it was considered hopeless to experimentally observe BEC
in an atomic gas with weak interactions. With the
development of techniques to trap and cool atoms, BEC was recently
observed directly in dilute atomic vapors\cite{and95, bra95, dav95}.
The new experimental achievements have stimulated great interest in
the theoretical study of inhomogeneous Bose gases. 

For BEC in the atom-trapping experiments, the critical temperature
$T_c$ can be altered by the spatially varying potential of trapping.
Furthermore, the noninteracting Bose gas can lead to unphysical
results. The interaction between atoms may have a significant
effect on the critical temperature. There have been several
investigations analyzing the dependence of the critical temperature
on the trapping potential and weak interaction in the Bose 
gas\cite{bag87,gio96}. Here we shall derive an analytical 
result for the critical temperature of a trapped interacting 
Bose gas by means of the local density approximation.

When the energy level spacing of the trapping potential is much
smaller than $k T = \beta^{-1}$ , the local density approximation
\cite{gol81,oli89,chou96} is a good approximation. Space may 
then be divided into
small cells, and in each cell we may consider the trapping
potential energy $V({\bf r})$ to be a constant. In this approximation
for an ideal Bose gas, the local density $\rho({\bf r})$ of
particles in the gaseous phase (i.e., without BEC) is
\begin{equation}
\rho ({\bf r})=\lambda ^{-3}\;g_{3/2}(\xi ) ,
\label{eq-lda1}
\end{equation}
where
\begin{equation}
\lambda = \left(\frac{2\pi \hbar ^2}{m k T}\right)^{1/2} ,
\label{wl}
\end{equation} 
\begin{equation}
\xi =z\exp[-\beta V({\bf r})], \qquad z = e^{\beta \mu}, 
\qquad g_\nu (x)=\sum_{j=1}^\infty j^{-\nu }x^j,
\label{def}
\end{equation}
and $\mu$ is the chemical potential. Equivalently, the density of
states at energy $\varepsilon$ is\cite{bag87}
\begin{equation}
\rho(\varepsilon) = \frac{2 \pi (2 m)^{3/2}}{h^3}
	\int \sqrt{\varepsilon - V({\bf r})} d^3 r .
\label{ed}
\end{equation}
We may now extend the argument of BEC for rigid walls\cite{huang87}
to the case of the spatially varying trapping potential $V({\bf r})$.
From Eq.~(\ref{eq-lda1}), the total number of particles can be written 
as 
\begin{equation}
N = N_0 + \int_V \rho({\bf r}) d^3 r ,
\label{tn}
\end{equation}
where $N_0$ is the number of particles in the ground state. Let
us consider a generic power-law potential as in Ref.~\cite{bag87}
\begin{equation}
V({\bf r})=\epsilon _1\left|\frac x{L_1}\right|^p+\epsilon _2\left|
\frac y{L_2}\right|^l+\epsilon_3\left|\frac z{L_3}\right|^q.
\label{pot} 
\end{equation}
Since $g_{3/2}(\xi)$ is a bounded, positive, monotonically 
increasing function of $\xi$, for the potential (\ref{pot}),
the integral in Eq.~(\ref{tn}) reaches its maximum value at 
fugacity $z=1$ or $\mu=0$ . From Eqs.~(\ref{def}) and (\ref{tn})
the critical temperature $T_c$, at which the ground
state begins to take on macroscopic values, is determined by
\begin{eqnarray}
N \lambda_c ^3 &=& \int d {\bf r} g_{3/2} [\exp(- \beta_c V({\bf r}) ) ] 
	\nonumber\\
	       &=& \sum_{j=1}^{\infty} j^{-3/2} 
	       \int \exp(- j\beta_c V({\bf r}))\; d {\bf r} .
\label{cond}	     
\label{n-it}
\end{eqnarray}
The definition of the Gamma function
\begin{equation}
\Gamma (z) = \int_0^{\infty} t^{z-1} e^{-t} dt
\end{equation}
implies
\begin{equation}
\int_0^{\infty} dx\; \exp(-a x^p) = \frac{a^{-1/p}}{p}\; \Gamma(1/p) .
\end{equation}
For the specific choice (\ref{pot}) of the potential, expression 
(\ref{cond}) then becomes
\begin{eqnarray}
N \lambda_c ^3 &=& \sum^{\infty}_{j=1} \frac{8 j^{-3/2}}{(j \beta_c)^{\eta-1/2}}
       \frac{L_1 L_2 L_3}{\epsilon_1^{1/p}\epsilon_2^{1/l} \epsilon_3^{1/q}}
       \frac{\Gamma(1/p)\Gamma(1/l)\Gamma(1/q)}{plq} \nonumber \\
               &=& 8\;\zeta (\eta +1)\;\frac{L_1 L_2 L_3\;I(p,l,q) }
               {\beta_c^{\eta -1/2}\;
       \epsilon_1^{1/p}\;\epsilon _2^{1/l}\;\epsilon _3^{1/q}} ,
\label{n-i-c}
\end{eqnarray}
where
\begin{equation}
I(p,l,q)=(plq)^{-1}\;\Gamma (1/p)\Gamma (1/l)\Gamma (1/q), 
\end{equation} 
and $\zeta (\nu )=\sum_{j=1}^\infty j^{-\nu }$ is the Riemann zeta 
function and $\eta =1/p+1/l+1/q+1/2$ . Noticing the dependence
(\ref{wl}) of the thermal wavelength $\lambda$ on the temperature,
we finally obtain
\begin{equation}
k T_c=\left(\frac{N\;(2\pi \hbar ^2)^{3/2}\;\epsilon _1^{1/p}\;\epsilon
_2^{1/l}\;\epsilon _3^{1/q}}{8\;m^{3/2}\;\zeta (\eta +1)\;
L_1 L_2 L_3\;I(p,l,q) }\right)^{1/(\eta+1)}. 
\label{i-t}
\end{equation}
From Eqs.~(\ref{tn}) and (\ref{n-i-c}) we can obtain the fraction of 
condensation at a temperature $T$ below $T_c$ 
\begin{equation}
\frac{N_0}{N} = 1 - \left(\frac{T}{T_c}\right)^{\eta + 1}
\label{cf-i}
\end{equation}
where $N_0$ is the number of Bose particles in the BEC state.

For the harmonic potential 
\begin{equation}
V({\bf r})=\frac{1}{2} m\omega _{\perp }^2r_{\perp }^2+\frac{1}{2} 
m\omega _z^2r_z^2 ,
\label{harmonic}
\end{equation}
the critical temperature and fraction of condensation reduce to
\begin{equation}
k T_c=\hbar\;\left(\frac{N\;\omega _{\perp }^2\;
\omega _z}{1.202}\right)^{1/3},
\label{t-har} 
\end{equation}
\begin{equation}
\frac{N_0}{N} = 1 - \left(\frac{T}{T_c}\right)^3 .
\label{fc-har}
\end{equation}
where we have made use of
\begin{equation}
\zeta (3)=1.202,\qquad I(2,2,2)= \pi^{3/2} /8.
\end{equation}
The results (\ref{t-har}) and (\ref{fc-har}) were first obtained 
in Ref.~\cite{gro50} and have been
verified in a recent experiment\cite{mew96}.

Relation (\ref{i-t}) for the generic power-law potential may
also be derived in terms of Eq.~(\ref{ed}). In Ref.~\cite{bag87}
it is given that
\begin{equation}
k T_c=\left[\frac{h^3}{2\pi (2m)^{3/2}}\frac N{L_1 L_2 L_3}\frac{\epsilon
_1^{1/p}\;\epsilon _2^{1/l}\;\epsilon _3^{1/q}}{F(p,l,q)Q(\eta )}\right]
^{1/(\eta +1)} 
\label{i-t-s}
\end{equation}
where 
\begin{equation}
Q(\eta )=\int_0^\infty \frac{\theta ^\eta }{\exp(\theta )-1}d\theta ,
\end{equation}
\begin{equation}
F(p,l,q)=8\;\left[\int_0^1 dx\;(1-x^p)^{1/2+1/q+1/l}\right] \left[\int_0^1 
dx\; (1-x^l)^{1/q+1/2}\right] \left[\int_0^1 dx\; (1-x^q)^{1/2}\right] .
\end{equation}
In fact, we may integrate both $Q$ and $F$ to obtain
\begin{equation}
Q(\eta) = \sum_{j=1}^{\infty} \int_0^{\infty} \theta^{\eta} 
	e^{-j\theta} d\theta = \zeta(\eta+1)\; \Gamma(\eta+1) ,
\end{equation}
\begin{equation}
F(p,l,q) = 4 \sqrt{\pi} \; I(p,l,q)/\Gamma(\eta+1) ,
\end{equation}
where in deriving the last equation we have used
\begin{equation}
\int^1_0 (1-x^p)^{\nu -1}dx = \frac{1}{p}\int^1_0 (1-t)^{\nu-1} 
	t^{\frac{1}{p} -1} dt = \frac{1}{p} B(1/p,\nu )
	=  \frac{\Gamma(\nu )\Gamma(1/p) }
	{p\;\Gamma(\nu +1/p) }.
\end{equation}
It can then be verified that Eq.~(\ref{i-t-s}) coincides with
(\ref{i-t}) as it should.

Along similar lines we may analyze the critical temperature of
a trapped interacting Bose gas. In Ref.~\cite{chou96} it has been 
given that when the $s$-wave scattering length $|a|\ll \lambda$,
in the local density approximation the local density of the gaseous 
phase as given in Eq.~(\ref{eq-lda1}) is replaced by 
\begin{equation}
\rho ({\bf r})=\lambda ^{-3}g_{3/2}(\tilde{\xi} ), 
\label{eq-lda2}
\end{equation}
with
\begin{equation}
\tilde{\xi} =z\exp[-\beta V({\bf r})-4a\lambda ^2\rho ({\bf r})], 
\label{xi-ni}
\end{equation}
where the term containing the scattering length $a$ describes
the mutual interaction of atoms.

From Eqs.~(\ref{eq-lda2}) and (\ref{xi-ni}) we have
\begin{equation}
\rho(0) = \lambda^{-3} g_{3/2}[z \exp(-4 a \lambda^2 \rho(0)].
\end{equation}
Thus, at the critical temperature the fugacity is
\begin{equation}
z_c = \exp[4 a \tilde{\lambda}_c^2 \rho_c(0)]
\end{equation}
where 
\begin{equation}
\rho_c(0) =\tilde{\lambda}_c^{-3} g_{3/2}(1) . 
\end{equation}
For this $z_c$ Eq.~(\ref{eq-lda2}) becomes
\begin{eqnarray}
\rho ({\bf r})\tilde{\lambda}_c ^3 &=& g_{3/2}[z_c \exp(-\tilde{\beta}_c V({\bf 
r}) 
	-4a \tilde{\lambda}_c ^2\rho ({\bf r}))] \nonumber \\
	&\approx& g_{3/2}[\exp(-\tilde{\beta}_c V({\bf r}))] 
	+4 a \tilde{\lambda}_c^2\; [\rho_c(0) - \rho({\bf r})] \;
	g_{1/2}[\exp(-\tilde{\beta}_c V({\bf r}))] \nonumber \\
	&\approx& g_{3/2}[\exp(-\tilde{\beta}_c V({\bf r}))] 
	+4 a \tilde{\lambda}_c^2 \rho_c(0) \;g_{1/2}[\exp(-\tilde{\beta}_c 
	V({\bf r}))]\nonumber \\
	&{}&\quad -4 a \tilde{\lambda}_c^{-1}\; g_{3/2}[\exp(-\tilde{\beta}_c 
	V({\bf r}))]\; g_{1/2}[\exp(-\tilde{\beta}_c V({\bf r}))] ,
\label{rho}
\end{eqnarray}
where we have used
\begin{equation}
z\frac{d}{dz}\;g_{3/2}(z)=g_{1/2}(z) , 
\end{equation}
and kept only the lowest order in $a/ \lambda$. The determination of
critical temperature $\tilde{T}_c$ for the interacting Bose gas involves 
integration of $\rho({\bf r})$
in (\ref{rho}). After some calculations similar to those
in the derivation of (\ref{n-i-c}) from (\ref{cond}), we find
\begin{equation}
N \tilde{\lambda}_c^3 = 8\; [\;\zeta(\eta+1) + 4 a 
\tilde{\lambda}_c^{-1}\; (\; \zeta(3/2) \zeta(\eta)
   -\chi(\eta-1/2)\;)\;]\;  \frac{L_1 L_2 L_3\; I(p,l,q)}
   {\beta_c^{\eta -1/2}\;\epsilon_1^{1/p}\;\epsilon _2^{1/l}\;\epsilon _3^{1/q}} 
,
\end{equation} 
where 
\begin{equation}
\chi (\nu )=\sum_{i,j=1}^\infty 
\frac{1}{i^{3/2}\;j^{1/2}}\;\frac{1}{(i+j)^\nu } .
\end{equation}
We finally obtain the critical temperature $\tilde T_c$
\begin{equation}
k \tilde{T}_c=\left(\frac{N\; (2\pi \hbar ^2)^{3/2}\;\epsilon _1^{1/p}\epsilon 
_2^{1/l}\epsilon
_3^{1/q}}{8\;m^{3/2} \;[\;\zeta (\eta +1)+4a\tilde{\lambda}_c ^{-1}\;(\;\zeta 
(3/2)\zeta (\eta )-\chi
(\eta -1/2)\;)\;]\;L_1 L_2 L_3\; I(p,l,q) }\right)^{1(\eta +1)}.
\label{ni-t}
\end{equation}
Comparing Eq.~(\ref{ni-t}) with Eq.~(\ref{i-t}) for the critical temperature
$T_c$ of the trapped ideal Bose gas, we can derive the relation
between the critical temperatures of BEC for ideal and nonideal Bose gases 
in the power-law potential
\begin{equation}
\frac{T_c}{\tilde{T}_c} = \left[1+ 4 a \tilde{\lambda}_c^{-1}\; \frac{\zeta(3/2) 
\zeta(\eta)
	- \chi(\eta-1/2)}{\zeta(\eta+1)} \right]^{1/(\eta+1)} ,
\end{equation}
which leads to
\begin{equation}
\tilde{T}_c = T_c - \frac{\zeta(3/2) \zeta(\eta)-\chi(\eta-1/2)}
	{(\eta+1)\;\zeta(\eta+1)}
	\left(\frac{8 km {T_c}^3}{ \pi \hbar^2} \right)^{1/2}\; a  .
\label{t_c-ni}
\end{equation}
From this analytical relation between the critical temperature and
the scattering length $a$, we see that the critical temperature $\tilde{T}_c$
decreases for $a>0$, and increases for $a<0$, compared with $T_c$ of
the ideal Bose gas. This agrees with Ref.~\cite{bag87}. 

We may also obtain an expression similar to the case of an ideal Bose gas 
for the condensate fraction  at a temperature $T<\tilde{T}_c$ to be
\begin{equation}
\frac{N_0}{N} = 1 - \left(\frac{T}{\tilde{T}_c}\right)^{\eta+1} + 4\; 
\frac{a}{\tilde\lambda_c}
	\frac{\zeta(3/2) \zeta(\eta)-\chi(\eta-1/2)}
	{\zeta(\eta+1)}\; \left(\frac{T}{\tilde T_c}\right)^{\eta+1}
	\;\left[1 - \left(\frac{T}{\tilde T_c}\right)^{1/2}\right]
\label{cf-ni}
\end{equation}
For the harmonic potential (\ref{harmonic}),   
Eq.~(\ref{cf-ni}) reduces to
\begin{equation}
\tilde T_c = T_c - \frac{\zeta(3/2) \zeta(2)-\chi(2/3)}
	{3 \;\zeta(3)}
	\;\left(\frac{8 km T_c^3}{ \pi \hbar^2} \right)^{1/2}\; a,
\end{equation}
which has been obtained in Ref.~\cite{gio96}.

The main effect of an external potential is to concentrate Bose 
particles. A repulsive mutual interaction 
between particles will weaken the effect of the external potential 
in concentrating particles around the center of the potential, hence lower 
the critical temperature. 

\bigskip
\begin{center}
\parbox{14cm}{
The authors thank Prof. Bai-lin Hao for encouraging and useful
discussions. This work was supported in part by the National
Natural Science Foundation of China.}
\end{center}

\end{document}